\documentclass[12pt]{article}
\usepackage{graphicx}
\usepackage{calc}
\usepackage{hyperref}
\hypersetup{breaklinks,colorlinks=true}
\usepackage{breakurl}
\newcommand{\dpr}{doi: \href{http://dx.doi.org/\dbi}{\nolinkurl{\dbi}}}
\newcommand{\upr}{url: \url{\URL}}

\newcommand\pubnumber{}
\newcommand\pubdate{}

\newcommand{\mathtf}{\mathrm}

\newcommand{\tev}{\mathtf{TeV}}
\newcommand{\gev}{\mathtf{GeV}}
\newcommand{\mev}{\mathtf{MeV}}
\newcommand{\fb}{\mathtf{fb}}

\newcommand{\nb}{\mathtf{nb}}

\newcommand{\ifb}{\fb^{-1}}

\newcommand{\lumi}{{\cal L}}
\newcommand{\br}  {{\cal B}}
\newcommand{\sqrs}{\sqrt{s}}
\newcommand{\eold}{\sqrs =    7~\tev}
\newcommand{\enew}{\sqrs =    8~\tev}

\newcommand{\lold}{\lumi \sim    5~\ifb}

\newcommand{\lnew}{\lumi \sim   20~\ifb}

\newcommand{\dm}{\Delta m}

\newcommand{\jpsi}{{J/\psi}}

\newcommand{\ups}[1]{{\Upsilon(#1\mathtf{S})}}
\newcommand{\bp}{B^+}
\newcommand{\bx}{B^\pm}

\newcommand{\mm}{\mu^+\mu^-}

\newcommand{\kk}{K^+K^-}

\newcommand{\jpmumu}{\jpsi \rightarrow \mm}

\newcommand{\brjpmm}{\br(\jpmumu)}

\newcommand{\bpjppk}{\bx \rightarrow \jpsi \phi K^\pm}

\newcommand{\jpc}{J^{PC}}

\newcommand{\dbi}{}
\newcommand{\URL}{}

\newcommand{\typeaut}{}

\newcommand{\typetit}{}
\newcommand{\typeref}{}

\newcommand{\typedoi}{}
\newcommand{\typeurl}{}

\def\padova{Dipartimento di Fisica e Astronomia\\
Universit\`a di Padova and INFN, I-35131 Padova, ITALY}

\def\Title#1{\begin{center} {\Large #1 } \end{center}}
\def\Author#1{\begin{center}{ \sc #1} \end{center}}
\def\Address#1{\begin{center}{ \it #1} \end{center}}

\newcommand\pubblock{\rightline{\begin{tabular}{l} \pubnumber\\
         \pubdate  \end{tabular}}}
\newenvironment{Abstract}{\begin{quotation}  }{\end{quotation}}
\newenvironment{Presented}{\begin{quotation} \begin{center} 
             PRESENTED AT\end{center}\bigskip 
      \begin{center}\begin{large}}{\end{large}\end{center} \end{quotation}}

\def\beq{\begin{equation}}
\def\eeq#1{\label{#1}\end{equation}}
\def\eeqn{\end{equation}}

\def\beqa{\begin{eqnarray}}
\def\eeqa#1{\label{#1}\end{eqnarray}}
\def\eeqan{\end{eqnarray}}

\let\bar=\overbar

\def\Dslash{\not{\hbox{\kern-4pt $D$}}}
\def\dslash{\not{\hbox{\kern-2pt $\del$}}}

\def\msb{{\bar{\ssstyle M \kern -1pt S}}}

\def\Journal#1#2#3#4{{#1} {\bf #2}, #3 (#4)}

\def\PLB{{\typeref Phys. Lett.}  B}
\def\PRL{\typeref Phys. Rev. Lett.}
\def\PRD{{\typeref Phys. Rev.} D}

\def\JHEP{\typeref J. High Energy Phys.}

\def\EPJC{{\typeref Eur. Phys. J.} C}

\begin{document}
\begin{titlepage}
\pubblock

\vfill
\Title{Exotic charmonium spectroscopy with CMS}
\vfill
\Author{Paolo Ronchese}
\Address{\padova}
\vfill
\begin{Abstract}
The latest results of CMS in the area of exotic quarkonium decays will be 
presented: observation of a peaking structure in $\jpsi\phi$ mass spectrum
in the decay $\bpjppk$, 
search for new bottomonium states in $\ups{1}\pi^+\pi^-$ mass spectrum, 
measurement of prompt $\jpsi$ pair production.
\end{Abstract}
\vfill
\begin{Presented}
Twelfth Conference on the Intersections \\
of Particle and Nuclear Physics\\
Vail, Colorado (USA), May 19-24, 2015\rule{0mm}{15pt}
\end{Presented}
\vfill
\end{titlepage}
\def\thefootnote{\fnsymbol{footnote}}
\setcounter{footnote}{0}

\section{Introduction}

The first, unexpected, charmonium state to be seen was the $X(3872)$, 
observed by Belle in 2003, in the decay 
$B^+ \rightarrow \jpsi K^+ \pi^+ \pi^-$~\cite{ref:xbell}. 
Several other states have been 
discovered since then, but their nature is not yet well understood. 
Several interpretations do exist, they can be tetraquarks, or hadronic 
molecules or hybrid mesons with a gluon content~\cite{ref:honia}. 
Their bottomonium partners are also looked for.

\section{Data samples and selections}

The results obtained from the analysis of data collected by CMS in 2011 
at $\eold$, corresponding to an integrated luminosity $\lold$, 
and in 2012 at $\enew$, corresponding to an integrated luminosity $\lnew$, 
will be shown in the following.

Dedicated triggers have been developed for the analyses to achieve a 
sustainable trigger rate when collecting data at the very high luminosities 
provided by LHC. The presence of two muons was required, with an invariant 
mass compatible with a $\jpsi$ or an $\Upsilon$, forming a secondary vertex 
displaced from the primary interaction point and a momentum direction 
compatible with the flight direction.

\section{Peaking structure in $\bpjppk$}

The $Y(4140)$ was first obserbved by CDF~\cite{ref:bpcdf}, with a 
mass near the $\jpsi \phi$ threshold and a narrow width: 
$\Gamma = (11.7^{+8.3}_{-5.0} \pm 3.7)~\mev$.
A consistent mass peak has been observed by D0~\cite{ref:bpkd0}, but has not 
been confirmed 
by Belle~\cite{ref:bpbel} and there's no evidence from LHCb~\cite{ref:bplhb}. 
Its mass is well above the threshold for open charm decay, and probably 
it's not a $P$-wave charmonium. Possible interpretations of its nature 
are a $D^*_s \bar{D}^*_s$ molecule with $\jpc=0^{++}$ or $\jpc=2^{++}$, 
or an hybrid charmonium with $\jpc=1^{-+}$; the hypothesis of 
a $c \bar{c} s \bar{s}$ tetraquark is compatible with $\jpc = 1^{++}$ but not 
$\jpc=0^{++}$~\cite{ref:pksjp}.

\begin{figure}[htb]
  \centering
  \includegraphics[trim=0pt 155pt 400pt 165pt, clip=true,height=115pt]{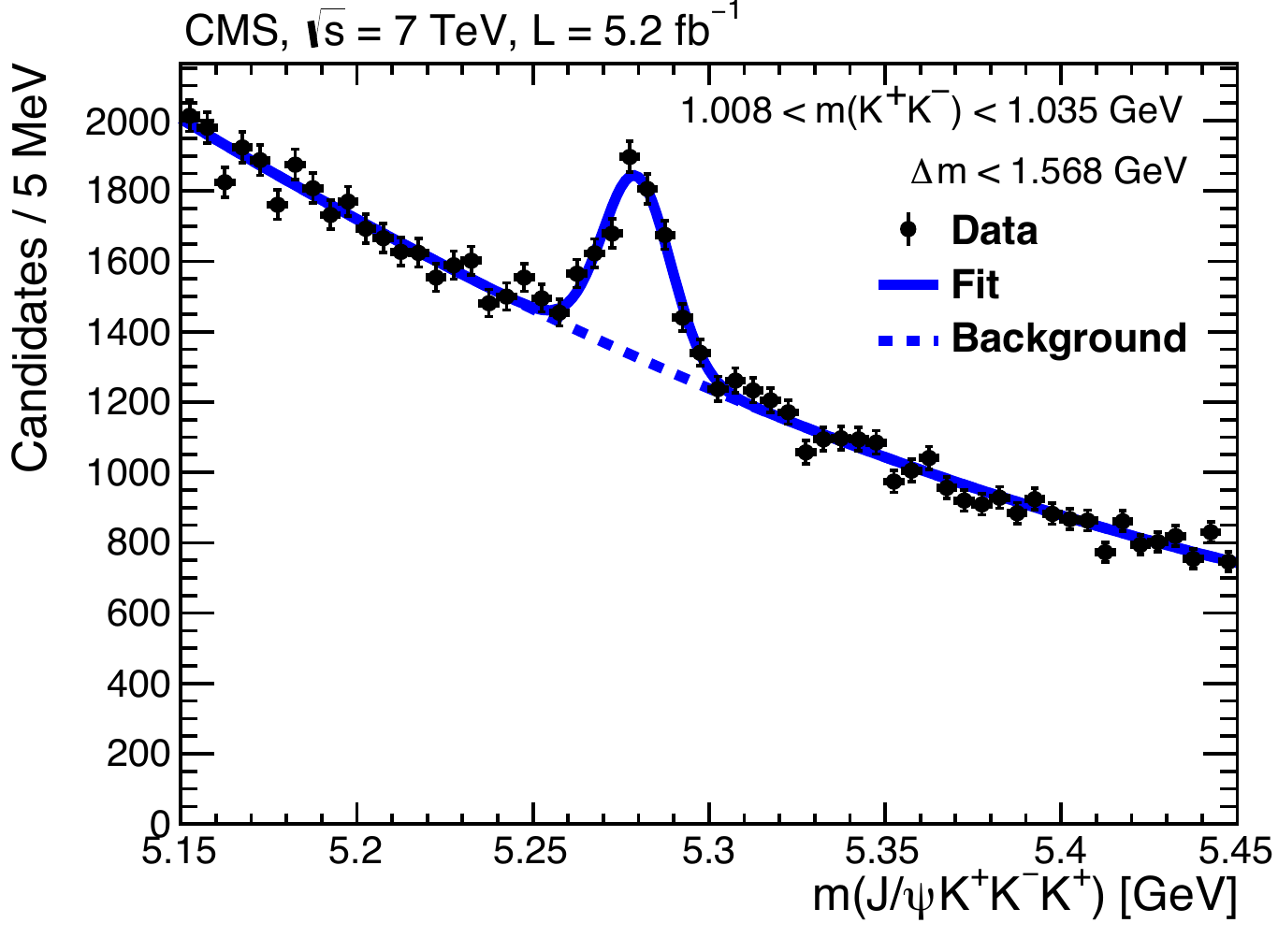}
  \includegraphics[height=115pt]{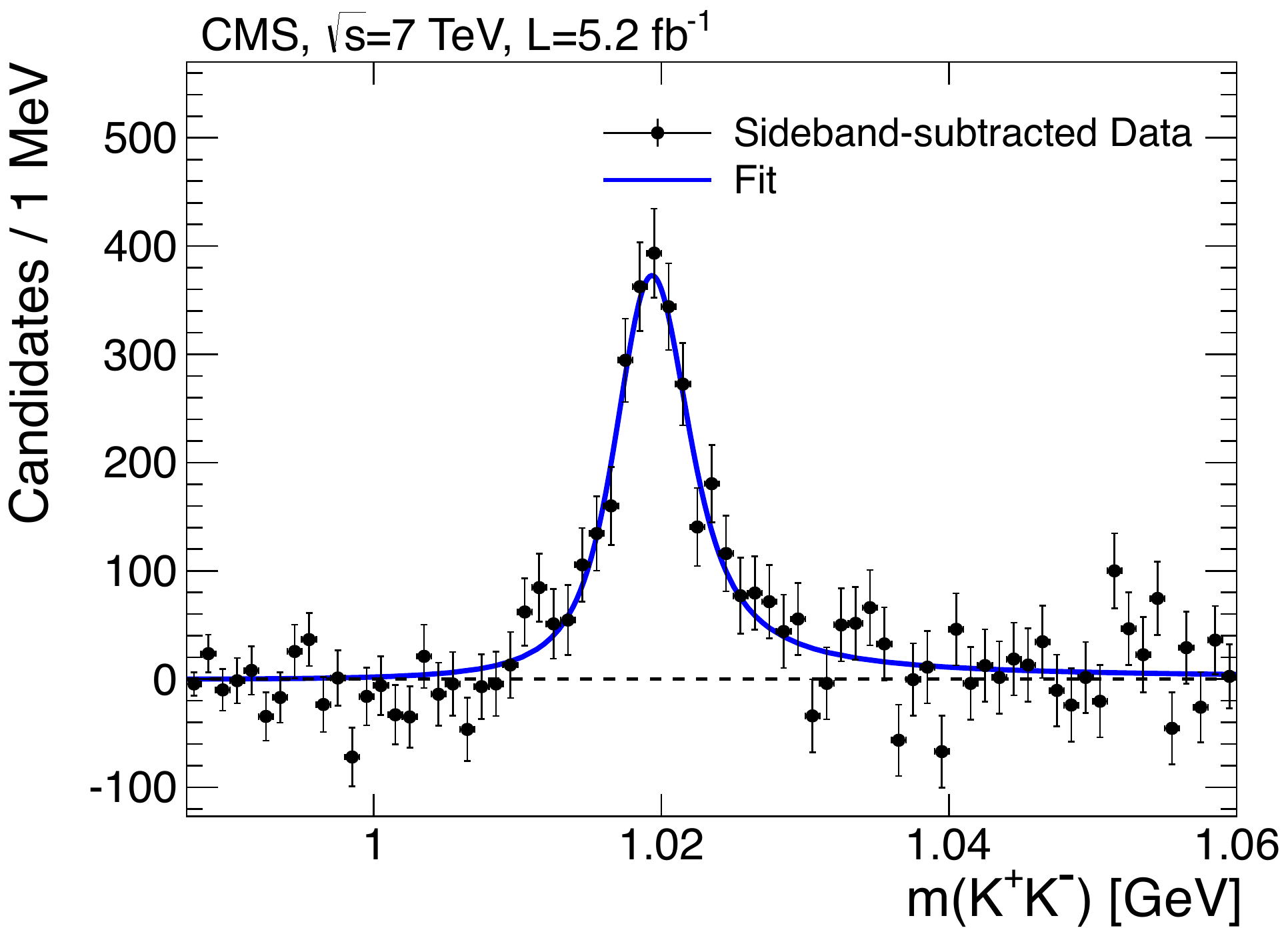}
  \caption{Invariant mass distribution for the $\jpsi \phi K^+$ (left) and 
    the $K^+ K^-$ (right).
    $K^+ K^-$ mass distribution has been obtained from candidates within 
    $\pm 3\sigma$ of the nominal $\bp$ mass and performing a sideband 
    subtraction.}
  \label{fig:mjphi}
\end{figure}

CMS reconstructed this state~\cite{ref:bpeak} combining $\jpsi$ candidates 
with 3 charged tracks with total charge $\pm1$ and consistent with 
originating from the $\jpsi$ vertex; a $K^+K^-$ pair with an invariant mass 
compatible with the $\phi$ was then looked for. 
Invariant mass distributions for the $\phi$ and the $\bp$ candidates 
are shown in fig.\ref{fig:mjphi}.

\begin{figure}[htb]
  \centering
  \includegraphics[height=115pt]{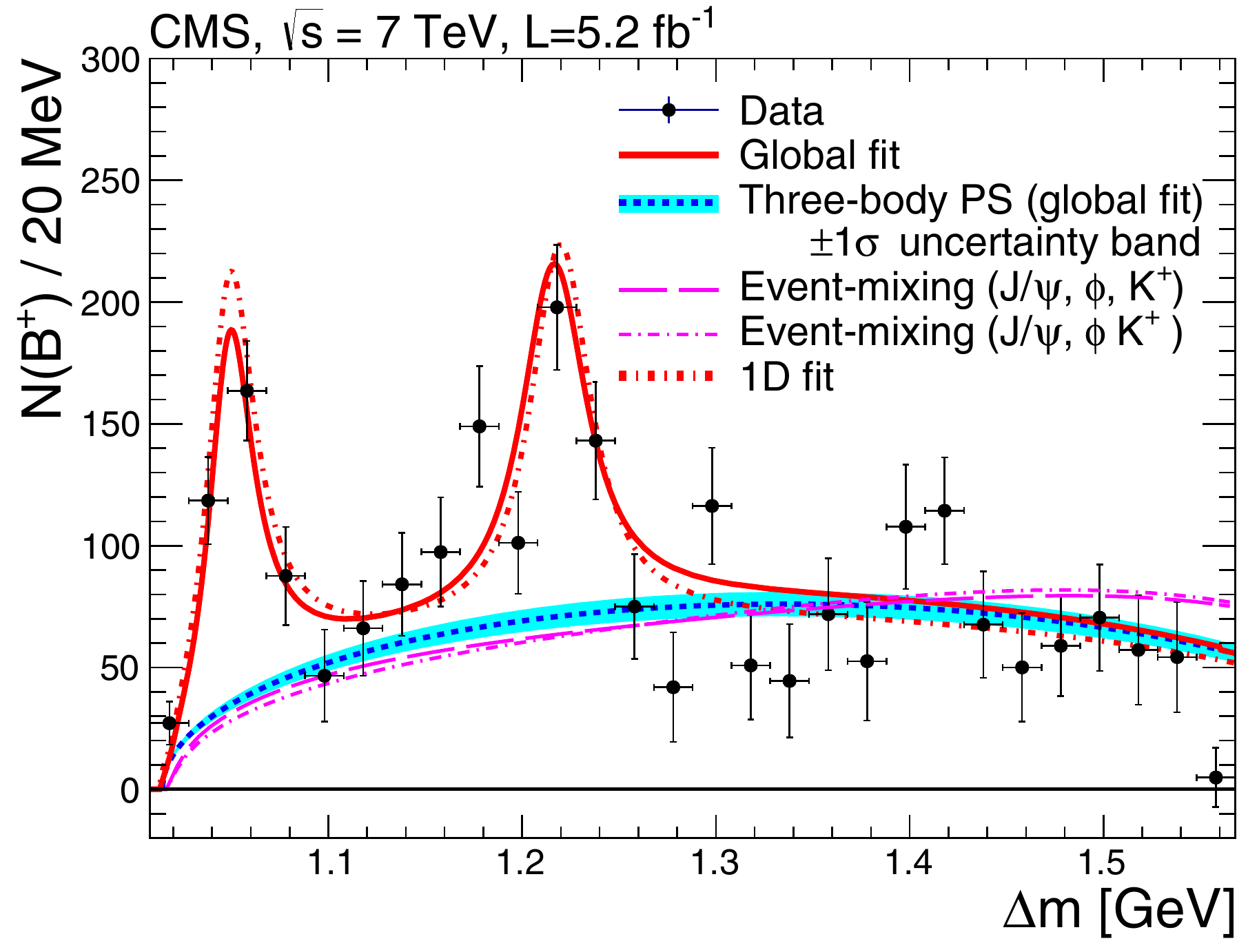}
  \caption{Number of $\bpjppk$ candidates as a function of $\dm$. The results
  of several fit procedures are 
  shown~\cite{ref:bpeak}.}
  \label{fig:dmfit}
\end{figure}

The $\jpsi \phi$ mass distribution is populated by a large combinatorial 
background, so the search is performed looking at the distribution 
of the mass difference 
$\dm \equiv m(\mm\kk) - m (\mm)$. 
Candidates have 
been divided into $20~\mev$ wide $\dm$ intervals, and for each interval 
the yield has been extracted fitting the total $\jpsi \phi K$ mass 
distribution. 
The signal has been fitted with two gaussians with mean fixed to $\bp$ mass 
and the background has been fitted with a second order polynomial. 
An unbinned maximum likelihood fit has been performed simultaneously 
for all intervals and the results are shown in fig.\ref{fig:dmfit}. 

The distribution of $\dm$ was then corrected for the detection and 
reconstruction efficiency, that was estimated with simulation. Different 
assumptions have been used for the polarization, and a negligible effect 
has been observed. The searched state has unknown quantum numbers, so 
different assumptions have been considered for the $\jpsi$ helicity angle 
distribution, and a 10\% effect was observed and included in the systematic 
uncertainty. 

\begin{figure}[htb]
  \centering
  \includegraphics[height=95pt]{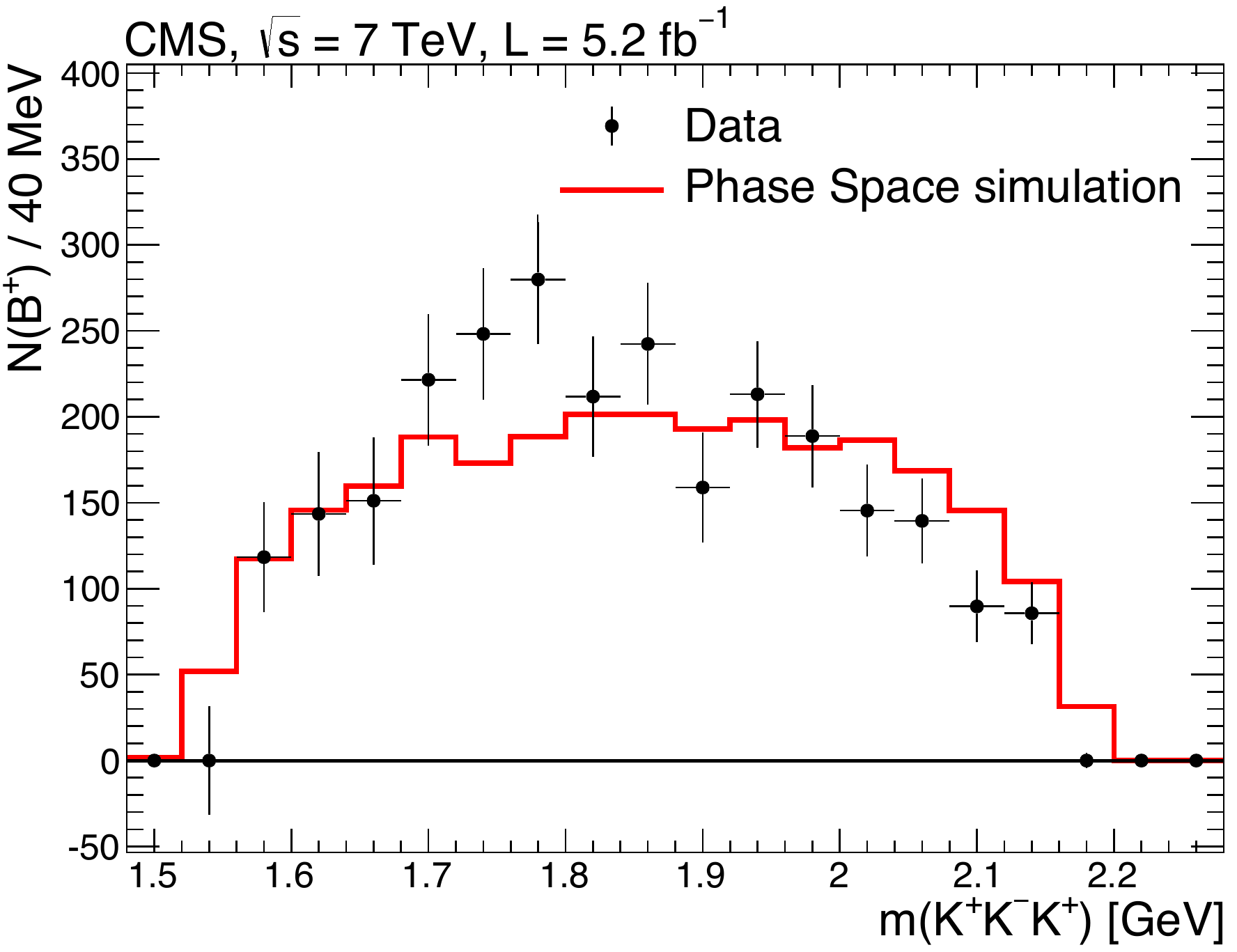}
  \includegraphics[height=95pt]{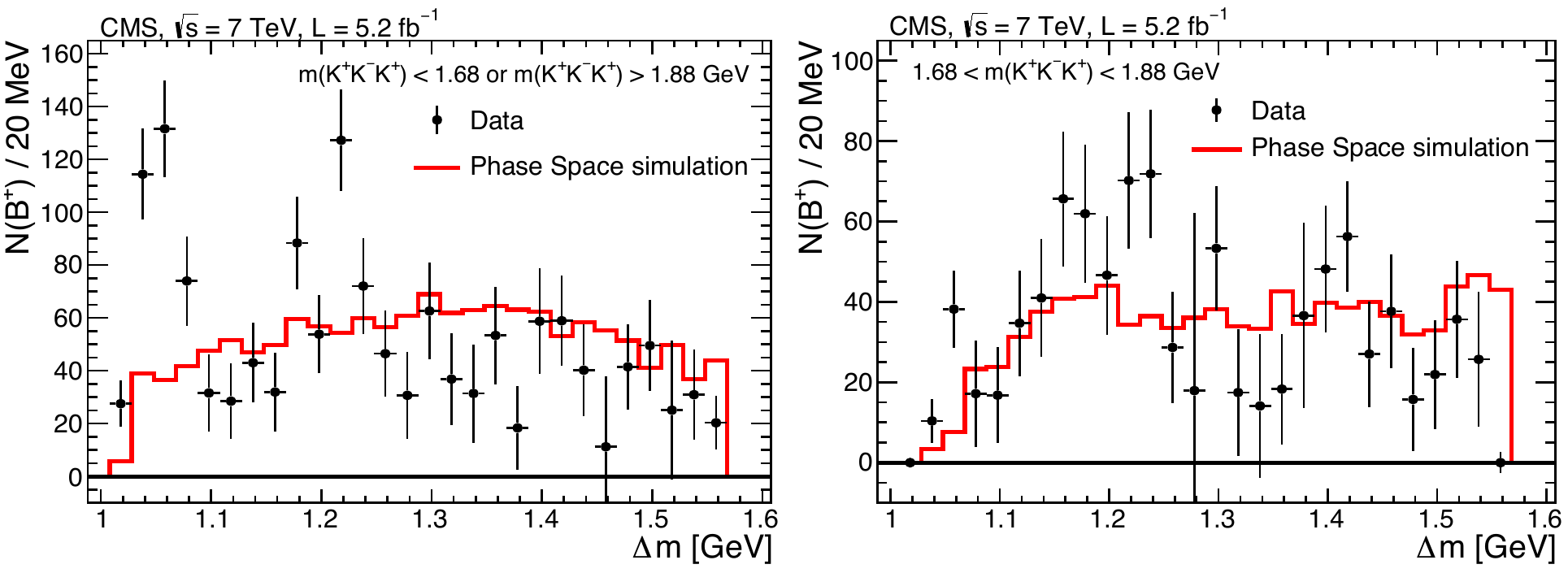}
  \caption{Yield of $\bp \rightarrow \jpsi K^+K^-K^+$ candidates as a function 
    of the $K^+K^-K^+$ invariant mass (left) and number of $\bpjppk$ candidates 
    as a function of $\dm$ requiring $m(\phi K)$ outside (middle) or inside 
    $[1.68,1.88]~\gev$ range (right).}
  \label{fig:dmchk}
\end{figure}

Checks have been performed to test the possibility that the observed 
peaks are actually reflections of resonances on the other 2-body systems, 
$\phi K$ or $\jpsi K$. Several resonances with various masses, widths and 
helicity angle distributions have been simulated, but none of them reproduced 
the observed spectrum. As an additional check, the mass distribution of the 
3 kaons has been investigated (fig.\ref{fig:dmchk}), and an excess in the 
region $1.7 \div 1.8~\gev$ is 
observed; 
the peaks are still visible also when the excess region is removed.

The analysis was also restricted to the region with $m(\phi K) > 1.9~\gev$ 
to remove possible effect from the $K_2(1770)$ and $K_2(1820)$ resonances, 
and the structure was still visible.
To compute systematic uncertainties and perform final cross checks 
different fit procedures and background estimations have been tried. 
The fit results for mass and width of the two peaks are reported in 
table~\ref{tab:jpres}.

\begin{table}[htb]
  \begin{center}
    \begin{tabular}{l|cccc}
           & yield         & $\dm~(\mev)$
           & $m~(\mev)$               & $\Gamma~(\mev)$        \\ \hline
      low  & $310 \pm  70$ & $1051.3 \pm 2.4$
           & $4148.0 \pm 2.4 \pm 6.3$ & $28^{+15}_{-11} \pm 19$  \\
      high & $418 \pm 170$ & $1217.1 \pm 5.3$
           & $4313.8 \pm 5.3 \pm 7.3$ & $38^{+30}_{-15} \pm 16$  \\ \hline
    \end{tabular}
  \end{center}
  \caption{Fit results for mass and width of the two peaks; $m_1$ and $m_2$ 
    values have been obtained adding the $\jpsi$ mass to the corresponding 
    $\dm$.}
  \label{tab:jpres}
\end{table}

\section{Search for a new bottomonium state decaying to $\ups{1} \pi^+\pi^-$}

The search for a bottomonium partner of $X(3872)$ has been conducted; 
such a state is predicted both by tetraquark model and hadronic molecular 
calculations; several mass predictions do exist, but 
a mass diference between the $X_b$ and the $\Upsilon$ larger than $1~\gev$ 
is predicted by all models. 
There are anyway important differences when 
comparing the decay $X_b \rightarrow \ups{1} \pi^+\pi^-$ with 
the decay of $X(3872) \rightarrow \jpsi \pi^+\pi^-$, leading to
a smaller isospin violation in $X_b$ decay than in $X(3872)$~\cite{ref:xbdif}. 

\begin{figure}[htb]
  \centering
  \includegraphics[height=95pt]{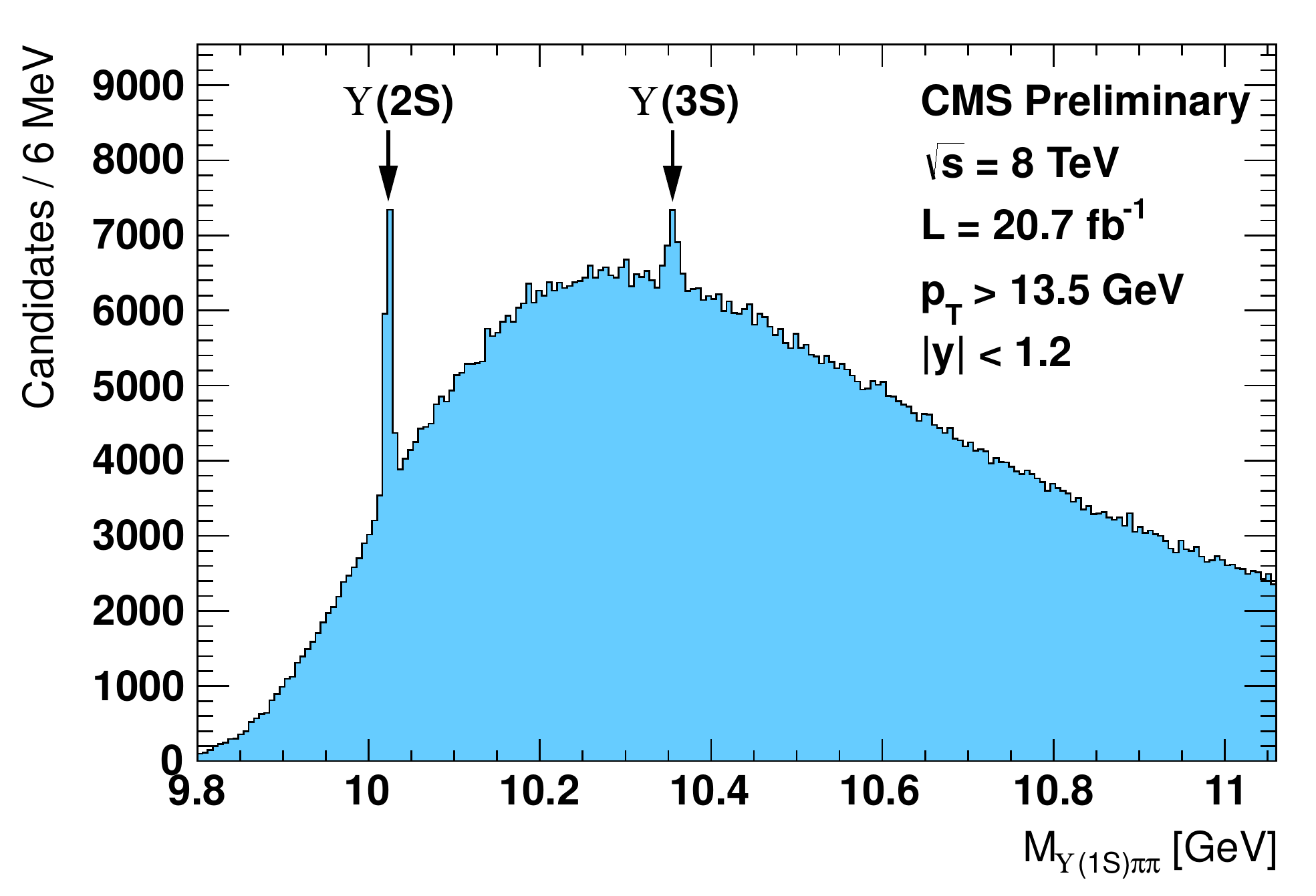}
  \includegraphics[height=95pt]{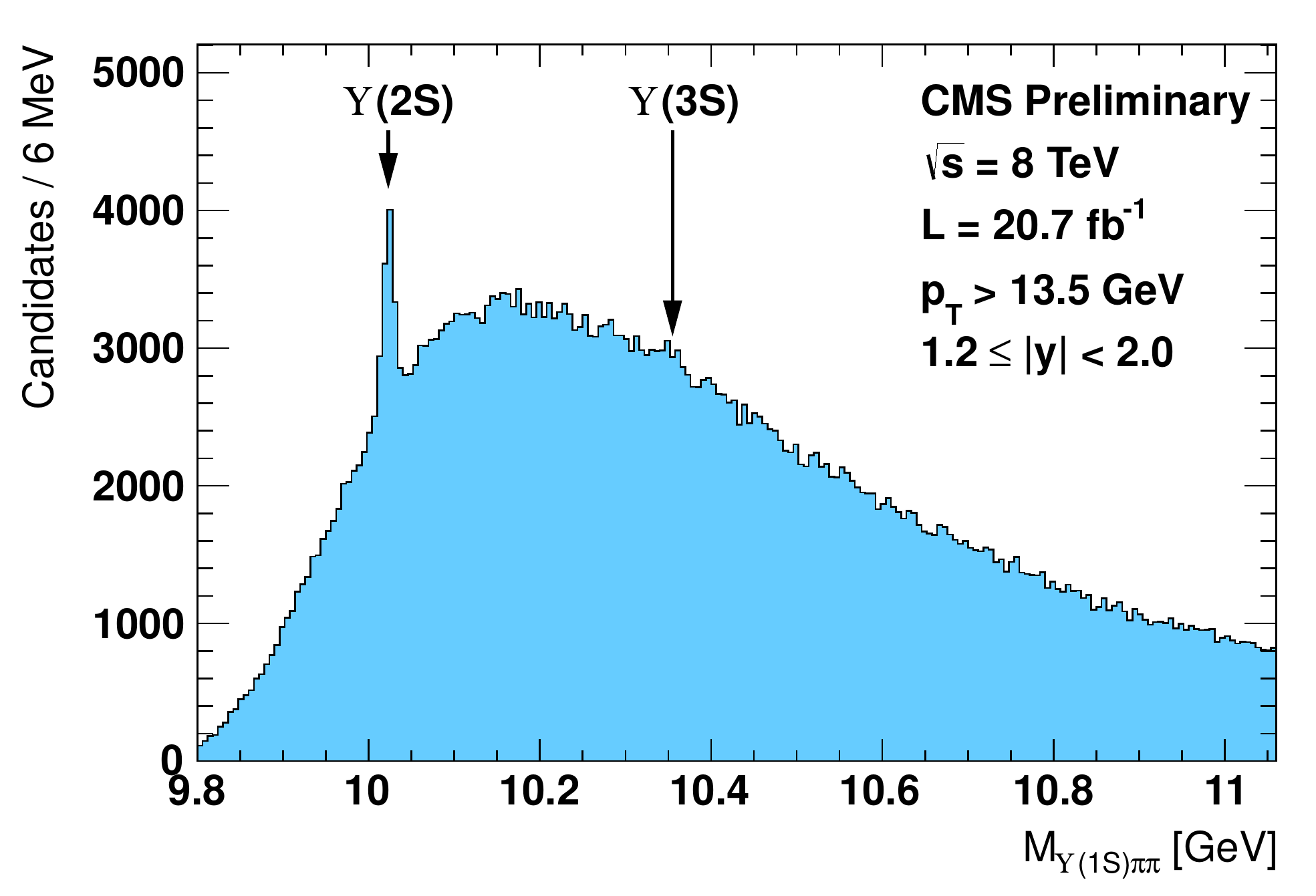}
  \caption{Reconstructed invariant-mass distributions of the candidates in 
    the barrel (left) and endcap (right) regions.}
  \label{fig:xbebm}
\end{figure}

CMS looked for this state~\cite{ref:xbsea} in events with a $\ups{1}$ candidate 
and 2 additional opposite-charged tracks, assumed to be pions. 
A common vertex has been fitted with 
muons and pions, and the invariant mass has been computed by constraining 
the dimuon mass to the $\ups{1}$. The mass distributions are shown 
in fig.\ref{fig:xbebm}, 
for rapidity intervals in the barrel and endcap regions, and the two peaks 
corresponding to $\ups{2}$ and $\ups{3}$ are visible.

\begin{figure}[htb]
  \centering
  \includegraphics[height=95pt]{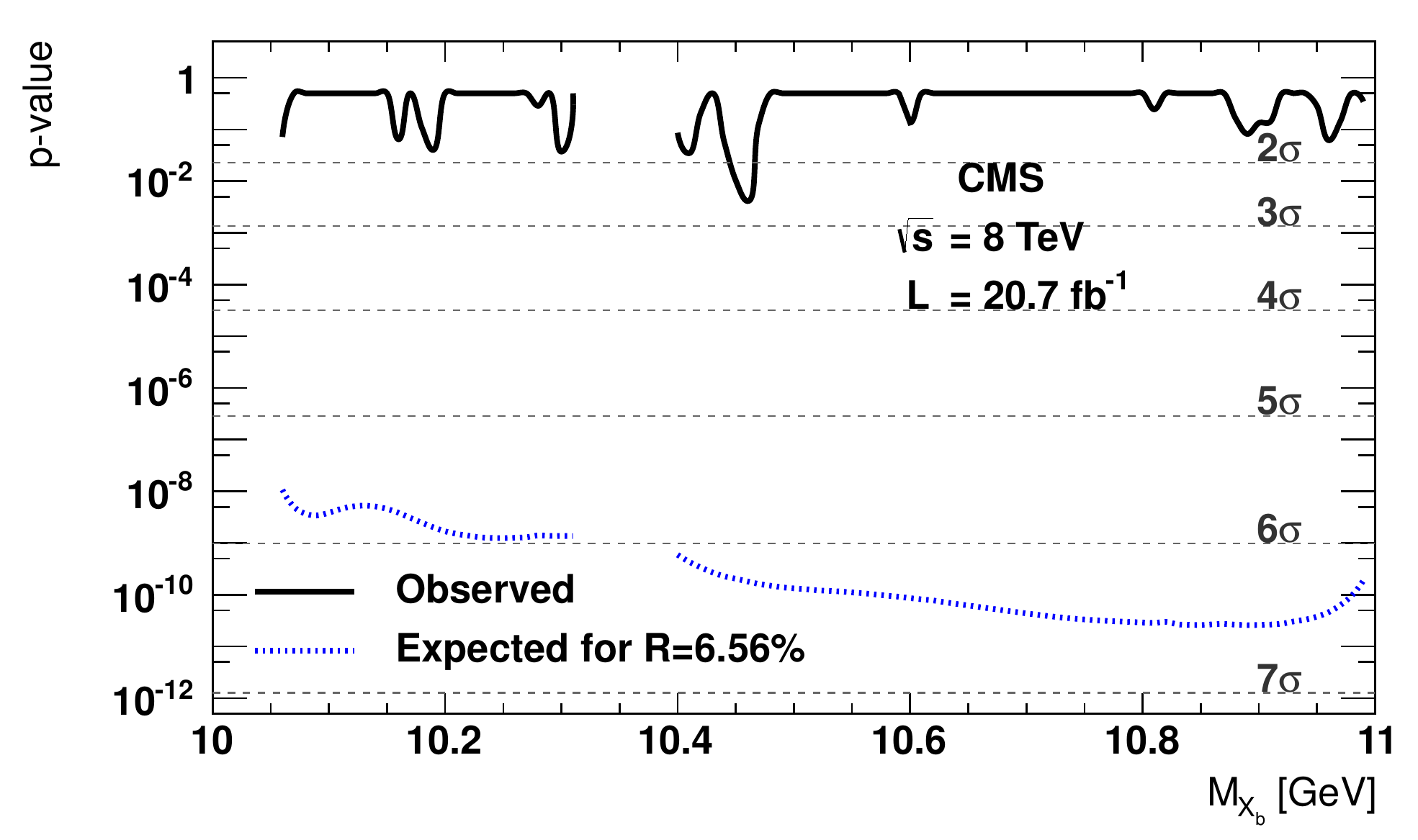}
  \includegraphics[height=95pt]{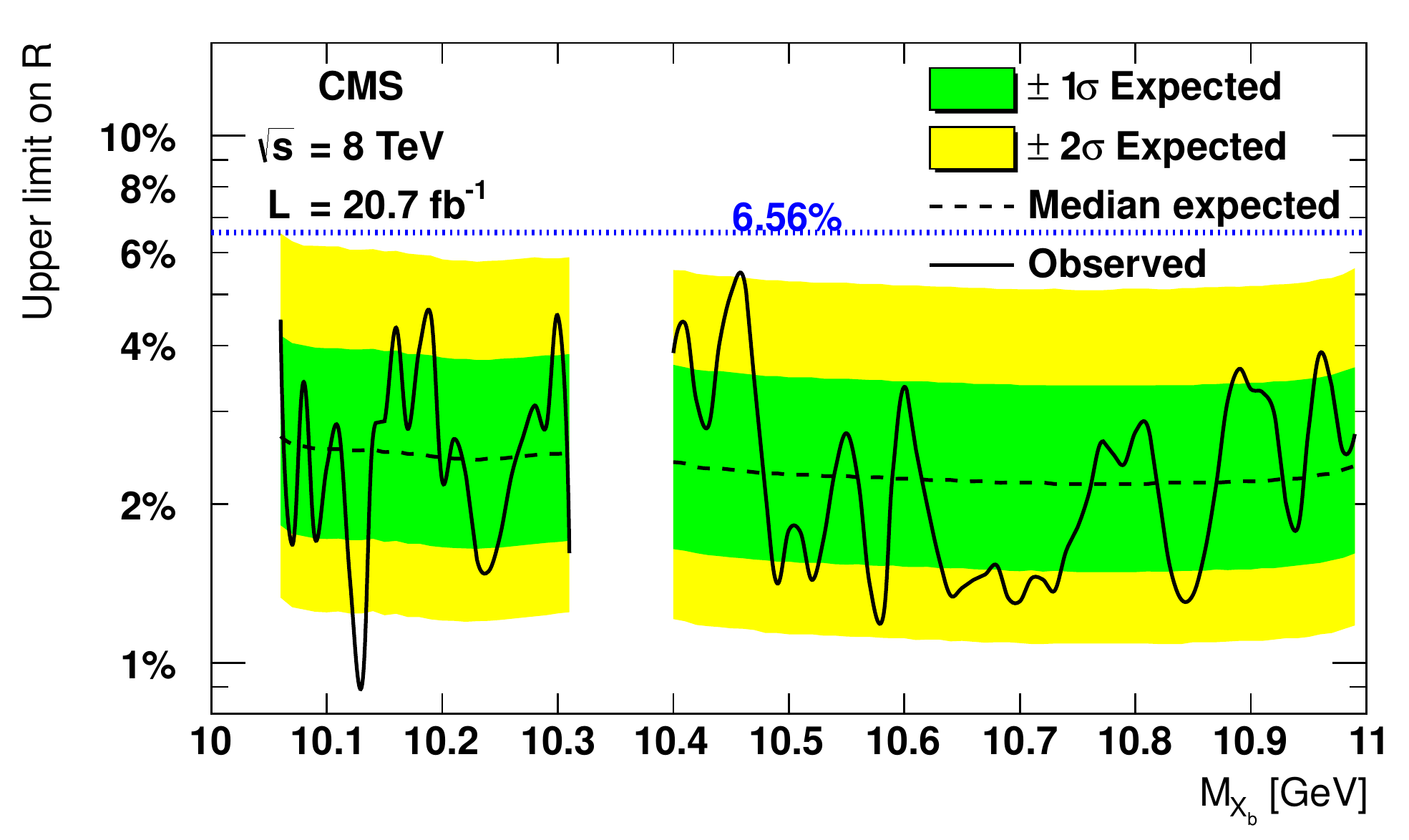}
  \caption{$P$-value (left) and upper limit at 95\% confidence level on $R$ 
    (right) as a function of the assumed $X_b$ mass. The lines at 6.56\% 
    correspond to the expectations for the analogous $X(3872)$ decay to
    $\jpsi \pi^+ \pi^-$.}
  \label{fig:xbres}
\end{figure}

The search has been performed estimating, for different hypotheses for 
the $X_b$ mass, the ratio $R$ of cross-sections for $X_b$ and $\ups{2}$:
\begin{displaymath}
R = \frac{\sigma(pp \rightarrow X_b     \rightarrow \ups{1} \pi^+\pi^-)}
         {\sigma(pp \rightarrow \ups{2} \rightarrow \ups{1} \pi^+\pi^-)}.
\end{displaymath}
The search has been performed in two mass regions, $[10.06,10.31]~\gev$ and \\
$[10.40,10.99]~\gev$, to exclude 
the $\ups{2}$ and $\ups{3}$; the mass spectrum has been fitted with a 
gaussian function for the resonances and a polynomial for background. 
The $X_b$ mass has been shifted by $10~\mev$ steps, while its width has 
been assumed to be small, and the resolution was taken from simulation. 
The ratio $R$ was then given by the ratio of the observed candidates scaled 
with the ratio of efficiencies. Signal strength, $P$-values and cross section 
limits have been computed versus the $X_b$ mass. Results are shown in 
fig.\ref{fig:xbres}.

Systematic uncertainties have been estimated by considering different 
models for dipion mass distribution and varying the efficiency dependence 
on $X_b$ mass. Different models for the mass distributions and resolutions 
have been considered as well as different assumptions for $X_b$ polarization.

\section{Measurement of prompt $\jpsi$ pair production}

In proton-proton collisions at LHC energies the very high parton density 
make possible having multiple parton scattering; single parton scattering 
is anyway assumed to dominate. The production of two $\jpsi$ is possible 
also in single parton scattering, but a pair of strongly correlated $\jpsi$ 
is expected, with a small rapidity difference $|\Delta y|$. Double parton 
scattering on the contrary allows multiple heavy-flavour production with 
large $|\Delta y|$~\cite{ref:djdif}. 
The production of $\jpsi$ pairs may undergo through the production of a 
color singlet, that is dominant at low transverse momenta, or through 
the production of a color octet turning into a singlet with the emission 
of a gluon, that becomes important at high $p_T$~\cite{ref:djcso}.
$\jpsi$ pairs can also be produced in the decay of $\eta_b$, but this decay 
is expected to be suppressed by non relativistic QCD~\cite{ref:djnrq}; 
other sources could be exotic states as tetraquarks~\cite{ref:djexo}. 

At CMS the differential cross-section was measured~\cite{ref:djpsi} 
versus 3 variables: the invariant mass $m_{\jpsi\jpsi}$ of the $\jpsi$ pair, 
the rapidity difference $|\Delta y|$ and 
the transverse total momentum $p_{T,\jpsi\jpsi}$. 
The differential cross-section is computed as 

\begin{displaymath}
  \frac{d\sigma(pp\rightarrow\jpsi\jpsi+X)}{dx}
  =\sum_i\frac{s_i}
  {a_i \cdot \epsilon_i \cdot (\brjpmm)^2 \cdot \Delta x \cdot \lumi}
\end{displaymath}

The sum is performed over all events~$i$ in an interval~$\Delta x$: 
$x$ is one of the 3 kinematical variables listed above, $s_i$ is the 
probability of the event to be signal, $a_i$ and $\epsilon_i$ are the 
acceptance and the detection efficiency, $\lumi$ is the integrated 
luminosity and $\br$ is the branching ratio for the $\jpsi$ to decay into 
two muons.

Events have been selected requiring at trigger level at least 
3~muons with different charge, and at reconstruction 4 muons with 
at least 3 of them matching the trigger; kinematic cuts are applied 
to transverse momenta, pseudorapidity of the single muons and 
rapidity of the $\jpsi$ candidates. Additional selections are applied 
cutting on the ``proper transverse decay length'' 
$ct_{xy} = (m_\jpsi/p_{T,\jpsi})\cdot L_{xy}$ and the ratio of the distance 
between the dimuon vertices $|\Delta\vec{r}|$ and its 
error $\sigma_{|\Delta\vec{r}|}$. 
Acceptance have been computed event by event generating a large number 
of events starting from measured $\jpsi$ 4-momenta and dividing the number 
of events surviving the acceptance cuts by the number of trials. 

The signal weights are obtained by a maximum likelihood fit of the 
distributions of the invariant masses of the two $\jpsi$, the proper 
tranverse decay length of the highest-$p_T$ $\jpsi$ and the distance 
significance. 
In the fit five categories have been considered: prompt $\jpsi$ pairs, 
that's the signal, events with at least one non-prompt $\jpsi$, events 
with one $\jpsi$ and a combinatorial dimuon, and events with two 
combinatorial dimuons. 

\begin{figure}[htb]
  \centering
  \includegraphics[height=145pt]{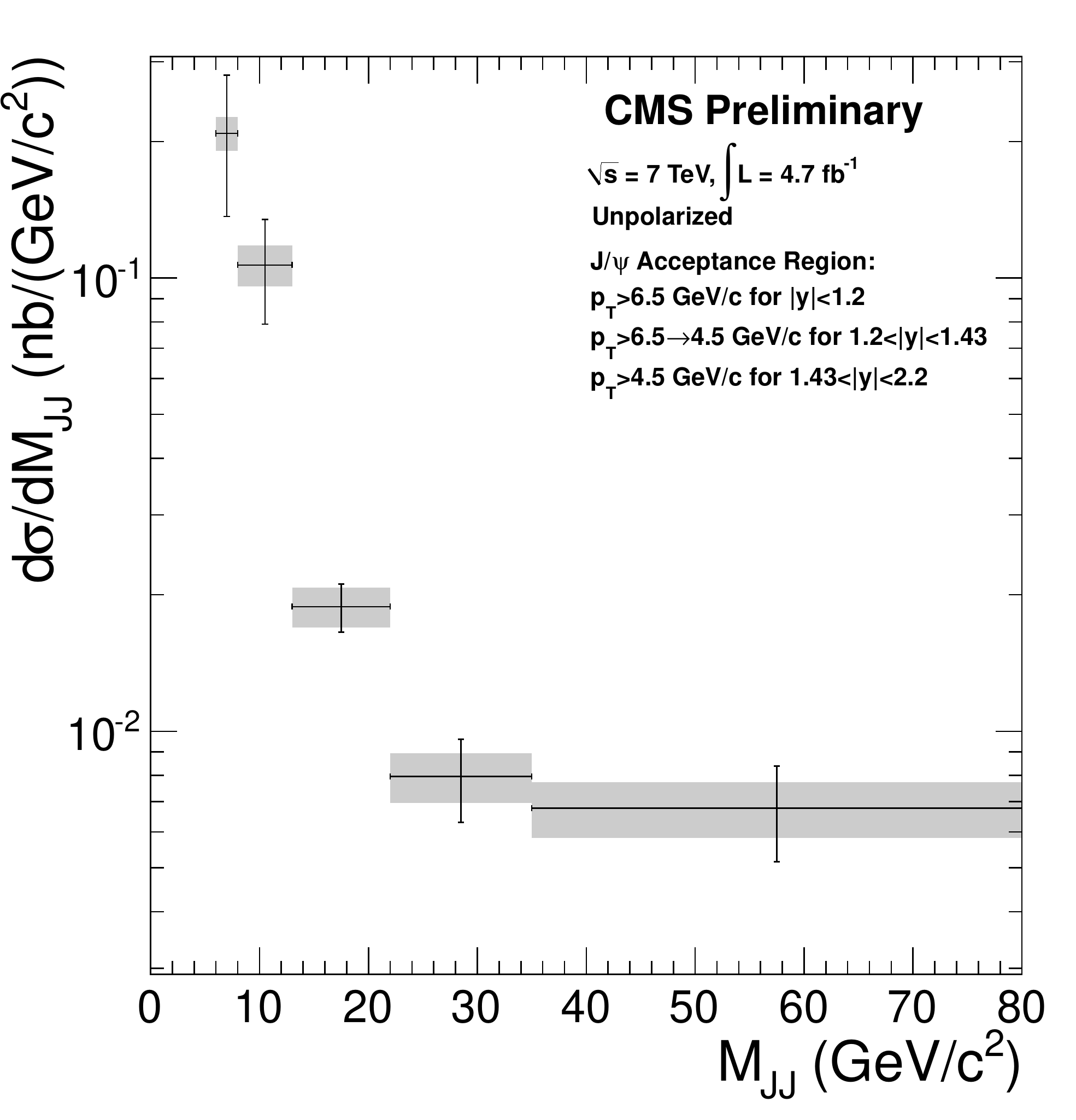}
  \includegraphics[height=145pt]{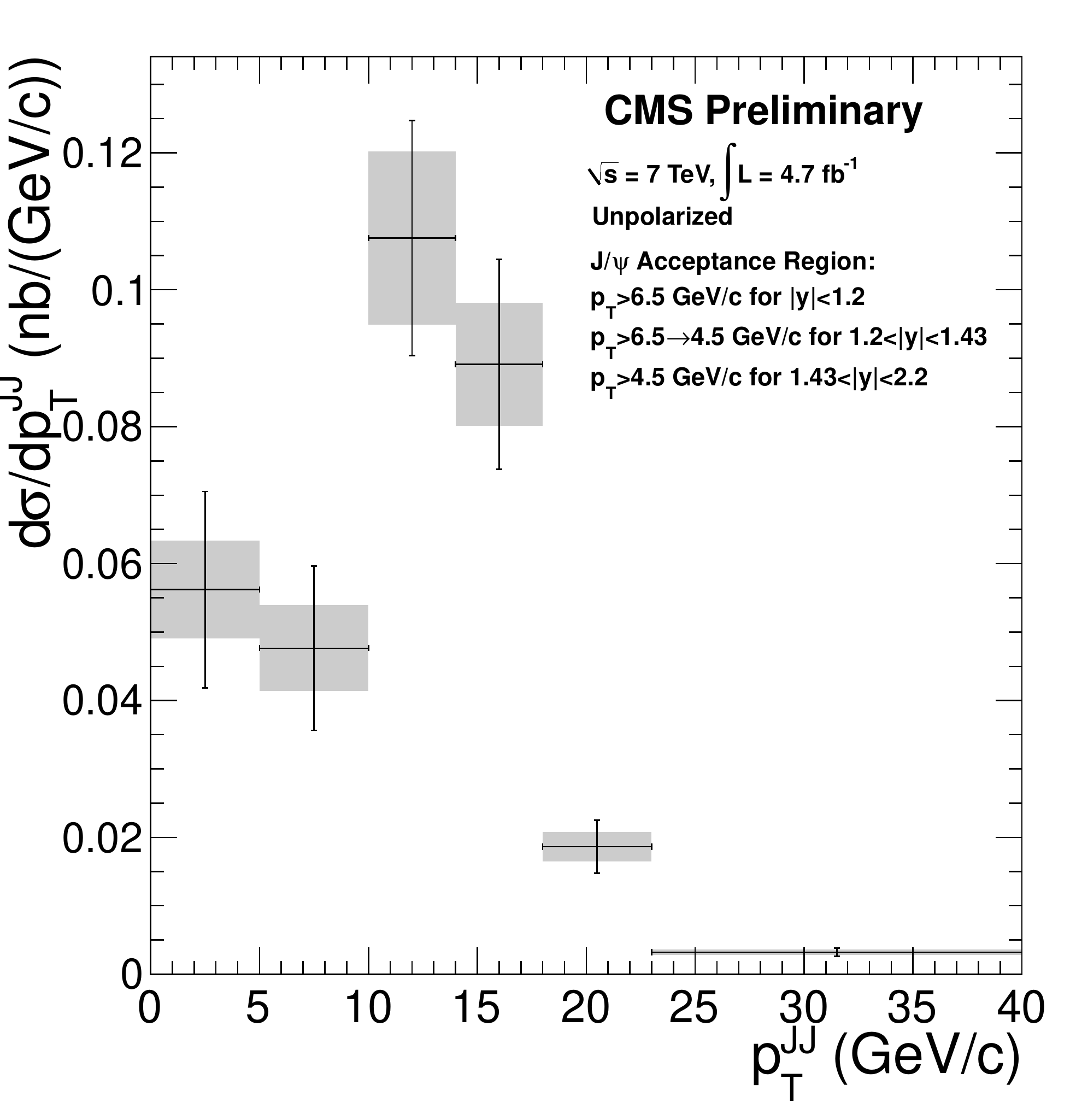}
  \includegraphics[height=145pt]{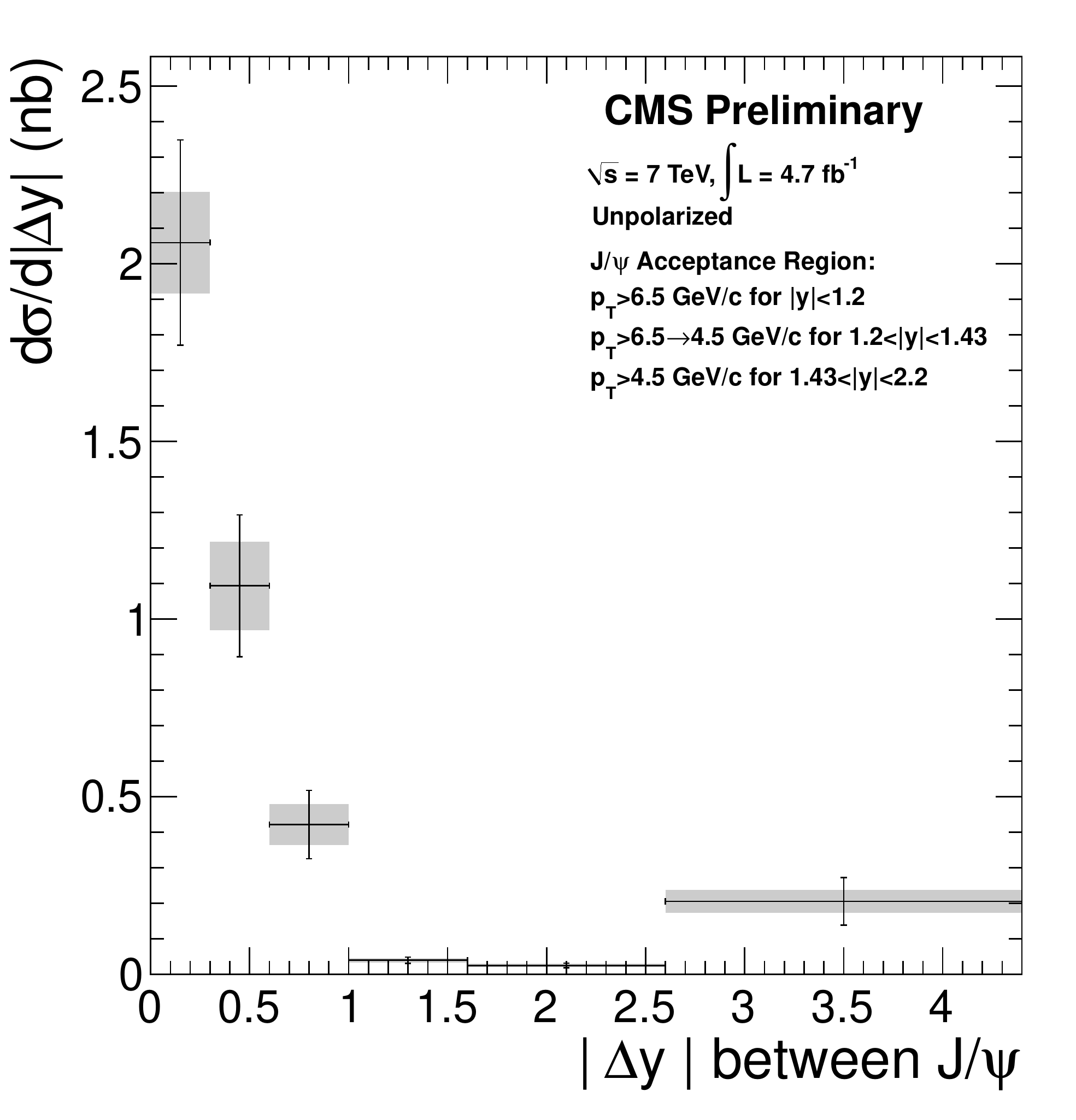}
  \caption{Differential cross-section for the production of a prompt 
    $\jpsi$ pair versus the invariant mass $m_{\jpsi\jpsi}$ of the $\jpsi$ 
    pair(left), the rapidity difference $|\Delta y|$ (middle) and the 
    transverse total momentum $p_{T,\jpsi\jpsi}$ (right).}
  \label{fig:djpsi}
\end{figure}

The differential cross-section versus $m_{\jpsi\jpsi}$, $|\Delta y|$ and 
$p_{T,\jpsi\jpsi}$ is shown in fig.\ref{fig:djpsi}; the total cross section is 

\begin{displaymath}
  \sigma(pp \rightarrow \jpsi \jpsi X) =
  (1.49 \pm 0.07 \pm 0.14)~\nb
\end{displaymath}

Looking at the $\eta_b$ mass region no excess is seen; the enhancement 
at large rapidity difference shows an hint of double parton scattering.

\section{Conclusions}

A structure has been observed in $\jpsi \phi$ mass in $\bpjppk$ decays, 
showing two peaks at masses $4148~\gev$ and $4313~\gev$.

A search has been conducted for a bottomonium partner of $X(3872)$ and a 
limit on the production of cross-section times the branching ratio 
to $\ups{1} \pi^+ \pi^-$ has been set.

A measurement of the cross-section for the production of a prompt 
$\jpsi$ pair has been performed, and hints for double parton scattering 
have been observed, while no evidence appeared of 
$\eta_b \rightarrow \jpsi \jpsi$ decay.


\urlstyle{same}

\newcommand{\cmscoll}{CMS Collaboration}
\newcommand{\cmslhcb}{CMS and LHCb Collaborations}
\newcommand{\atlcoll}{ATLAS Collaboration}
\newcommand{\lhbcoll}{LHCb Collaboration}

\newcommand{\eal}{{\it et al.}}

\newcommand{\tit}{}

\newcommand{\tpr}{\tit,}

\end{document}